\documentclass[english,aip,journal=jcp,reprint]{revtex4-1}
\usepackage[T1]{fontenc}
\usepackage[utf8]{inputenc}
\usepackage{refstyle}
\usepackage{graphicx}
\usepackage[version=4]{mhchem}
\usepackage{microtype}
\usepackage[unicode=true,pdfusetitle,
 bookmarks=true,bookmarksnumbered=false,bookmarksopen=false,
 breaklinks=false,pdfborder={0 0 1},backref=false,colorlinks=false]
 {hyperref}

\makeatletter

\AtBeginDocument{\providecommand\eqref[1]{\ref{eq:#1}}}
\AtBeginDocument{}
\AtBeginDocument{}

\RS@ifundefined{subsecref}
  {\newref{subsec}{name = \RSsectxt}}
  {}
\RS@ifundefined{thmref}
  {\def\RSthmtxt{theorem~}\newref{thm}{name = \RSthmtxt}}
  {}
\RS@ifundefined{lemref}
  {\def\RSlemtxt{lemma~}\newref{lem}{name = \RSlemtxt}}
  {}

\makeatother

\begin{document}

\setcounter{topnumber}{3}
\setcounter{bottomnumber}{3}
\setcounter{totalnumber}{6}

\title{Reply to Comment on ``Efficient implementation of the
  superposition of atomic potentials initial guess for electronic
  structure calculations in Gaussian basis sets''}
\author{Susi Lehtola}
\email{susi.lehtola@alumni.helsinki.fi}
\affiliation{Department of Chemistry, University of Helsinki, P.O. Box
  55 (A. I. Virtasen aukio 1), FI-00014 Helsinki, Finland.}
\author{Lucas Visscher}
\email{l.visscher@vu.nl}
\affiliation{Division
  of Theoretical Chemistry, Vrije Universiteit Amsterdam,De Boelelaan
  1083, 1081 HV Amsterdam, The Netherlands.}
\author{Eberhard Engel}
\affiliation{Center for Scientific Computing, J. W. Goethe-Universität
  Frankfurt, Max-von-Laue-Strasse 1, D-60438 Frankfurt am Main,
  Germany.}

\begin{abstract}
  We reply to the Comment by Surjuse, Deng, Asadchev, and Valeev
  [arXiv:2603.16989] on our article ``Efficient implementation of the
  superposition of atomic potentials initial guess for electronic
  structure calculations in Gaussian basis sets'' [J. Chem. Phys. 152,
    144105 (2020)]. We do not disagree with their work, but point out
  that they appear to have misinterpreted a statement in our original
  abstract: our suggestion to evaluate the matrix elements of the
  superposition of atomic potentials (SAP) using two-electron integral
  code was made on grounds of portability rather than as a
  recommendation against a dedicated one-electron implementation. Such
  an implementation is in fact already available in existing
  codes---in {\sc Dirac}, for example, where a Gaussian SAP potential
  is formally indistinguishable from a Gaussian finite-nucleus model
  at the level of the nuclear attraction integral routine. We further
  note that both routes reduce to the same three-Gaussian overlap
  kernel through the Obara--Saika Laplace transform, and that SAP
  itself enjoys a rigorous theoretical justification through
  Theophilou's theorem on spherical densities and its extensions by
  Nagy.
\end{abstract}
\maketitle
\global\long\def\ERI#1#2{(#1|#2)}%
\global\long\def\bra#1{\Bra{#1}}%
\global\long\def\ket#1{\Ket{#1}}%
\global\long\def\erf#1{\text{\,erf\,}(#1)}%
\global\long\def\erfc#1{\text{\,erfc\,}(#1)}%

\newcommand*\ie{\emph{i.e.}}
\newcommand*\eg{\emph{e.g.}}
\newcommand*\citeref[1]{ref. \citenum{#1}}
\newcommand*\citerefs[1]{refs. \citenum{#1}}
\newcommand*\HelFEM{{\sc HelFEM}}
\newcommand*\Erkale{{\sc Erkale}}
\newcommand*\Gaussian{{\sc Gaussian}}
\newcommand*\Grasp{{\sc Grasp}}
\newcommand*\Dirac{{\sc Dirac}}
\newcommand*\PsiFour{{\sc Psi4}}
\newcommand*\Mathematica{{\sc Mathematica}}

In their comment\cite{Surjuse2026__} on our
work\cite{Lehtola2020_JCP_144105}, \citeauthor{Surjuse2026__} show
that the matrix elements of the Gaussian-fitted superposition of
atomic potentials (SAP)\cite{Lehtola2019_JCTC_1593} can be evaluated
by a modification of the Boys route\cite{Boys1950_PRSAMPES_542} for
one-electron integrals. We agree, and the reported speedups are
promising for machine-learning applications, which were first
explored, to the best of our knowledge, by
\citet{Fabrizio2022_DD_286}. We note, however, that the approach is
rather straightforward, and that it is already largely implemented in
existing codes. In fact, we already indicated this in the text of our original paper\cite{Lehtola2020_JCP_144105}: "\textit{We would especially like to point out that the implementation of the fitted SAP guess is fully analogous to the computation of the nuclear attraction matrix elements for finite nuclei with Gaussian distributions that is already available in several program packages.}" 

Our original remark, made in the abstract of reference \citenum{Lehtola2020_JCP_144105}, that the SAP matrix
elements could be evaluated using two-electron integral code reflects
the fact that both routes ultimately reduce to the same kernel. In the
Obara--Saika scheme,\cite{Obara1986_JCP_3963} the Coulomb kernel
in any Gaussian integral is handled through the Laplace transform as
\begin{equation}
  \frac{1}{r_{12}} = \frac{2}{\sqrt{\pi}} \int_0^{\infty}
  e^{-(\mathbf{r}_1-\mathbf{r}_2)^2 t^2} \, \mathrm{d}t,
  \label{eq:laplace}
\end{equation}
which turns a nuclear attraction integral into a three-Gaussian
overlap integral supplemented by a one-dimensional integration over
$t$, which eventually leads to the Boys function. A two-electron
integral also reduces to the same form: the Gaussian product theorem
collapses the bra and ket pairs each into a single Gaussian, so that
the four-center integral becomes a three-Gaussian overlap with an
additional $t$ integration as well. Because such two-electron routines
are universally available in all Gaussian-basis quantum chemistry
programs, we suggested the two-electron route as a portable
option.

However, structurally, the SAP matrix elements are nuclear attraction
integrals in which the point nucleus is replaced by a finite Gaussian
charge distribution; any program supporting a Gaussian nuclear model
therefore already has the required machinery. The combination with a
finite nuclear model is then immediate: a Gaussian nucleus and a
Gaussian SAP potential are formally indistinguishable at the level of
the integral routine, so the two Gaussian sets can simply be
concatenated and passed through the same call. In fact, one of us (LV)
implemented the SAP potential in \Dirac{}\cite{Saue2020_JCP_204104} in
2016 by calling the existing nuclear attraction routine
(based on Helgaker's implementation of the McMurchie--Davidson
scheme\cite{McMurchie1978_JCP_218,Helgaker2000__}) once per primitive
Gaussian and then summing over the exponents.  This capability has
therefore been available in \Dirac{} from the outset. The scheme used
handles the contraction at a high level rather than inside the
integral kernel; a low-level treatment as in \citet{Surjuse2026__}
would be more efficient, but since the SAP potential is built in DIRAC
only once at the start of a self-consistent field calculation, and the
cost of this construction is much smaller than a single Fock matrix
build, the construction cost is negligible overall and the speed is
not a practical concern. The same observation also applies to the use
of the portable route to implement the fitted SAP guess in terms of
two-electron integrals.

The dedicated one-electron implementation built by
\citet{Surjuse2026__} has the advantage that several terms in the
recursion relations can be eliminated, as originally pointed out for
density-fitting integrals by \citet{Ahlrichs2006_PCCP_7}, and we are
pleased to see this route spelled out and benchmarked in detail.

We finally note that the spherically averaged atomic potentials
underlying SAP have a rigorous theoretical justification.
\citet{Theophilou2018_JCP_74104} proved that the set of spherical
parts of the electron density around each nucleus uniquely determines
the external potential in atoms, molecules, and solids; the theorem
has undergone several extensions by Nagy\cite{Nagy2018_JCP_204112,
  Nagy2020_JPCA_148, Nagy2021_JCP_74103, Nagy2021_JCP_144108,
  Nagy2022_LMP_107, Nagy2023_C_119, Nagy2024_JCP_44120}.  This means
that the induced Kohn--Sham potential is spherically symmetric around
each nucleus, and inherits the full symmetry of the external
potential. The SAP construction may thus be regarded as an explicit
realization of this structure at the level of the initial guess. This
connection suggests that further developments along a physically
motivated basis for machine-learned density functionals---exploiting
the atom-centered spherical structure guaranteed by Theophilou's
theorem---are to be expected, and here the 2--5 $\times$ speedups
reported by \citet{Surjuse2026__} will be useful.

\bibliography{citations, citations_luuk}

\end{document}